\documentstyle[preprint,revtex]{aps}
\begin{document}
\draft
\hfill{ITD 92/93--11}\par
\begin{title}
Level-Spacing Distributions and the Airy Kernel
\end{title}
\author{Craig A.~Tracy\footnotemark[1]}
\footnotetext[1]{e-mail address: catracy@ucdavis.edu}
\begin{instit}
Department of Mathematics and Institute of Theoretical Dynamics,\\
University of California,
Davis, CA 95616, USA
\end{instit}
\author{Harold Widom\footnotemark[2]}
\footnotetext[2]{e-mail address: widom@cats.ucsc.edu}
\begin{instit}
Department of Mathematics,\\
University of California,
Santa Cruz, CA 95064, USA
\end{instit}
\begin{abstract}
Scaling level-spacing distribution functions in the ``bulk of the
spectrum'' in random matrix models of $N\times N$ hermitian matrices
and then going to the limit $N\rightarrow\infty$, leads to the
Fredholm determinant of the {\it sine kernel\/} $\sin\pi(x-y)/\pi (x-y)$.
Similarly a double scaling limit at the ``edge of the spectrum''
leads to the {\it Airy kernel\/} $\left[ {\rm Ai}(x) {\rm Ai}^\prime(y)
-{\rm Ai}^\prime(x) {\rm Ai}(y)\right]/(x-y)$.  We announce
analogies for this Airy kernel of the following properties of
the sine kernel: the completely integrable system of P.D.E.'s
found by Jimbo, Miwa, M{\^o}ri and Sato; the expression, in the
case of a single interval, of the Fredholm determinant in terms
of a Painlev{\'e} transcendent; the existence of a commuting
differential operator; and the fact that this operator can be
used in the derivation of asymptotics, for general $n$, of the
probability that an interval contains precisely $n$ eigenvalues.

\end{abstract}
\vfill\eject
\section{Introduction}
\label{sec:intro}
In this note we announce  new results
for the {\it level spacing distribution functions\/}
obtained from scaling
 random matrix models of $N\times N$ hermitian matrices
at the edge of the support of the (tree-level) eigenvalue
densities when the parameters of the potential $V$
are  not ``finely tuned.''  This universality class is already
present in the Gaussian Unitary Ensemble.
It is known \cite{mehta_book,bowick_etal} that these
distribution functions are expressible in terms of a Fredholm
determinant of an integral operator $K$ whose kernel involves
 Airy functions.
\par
  There are striking analogies between the
properties of this   {\it Airy kernel\/}
\[ K(x,y) = {A(x) A^\prime(y) - A^\prime(x) A(y) \over x-y}
\]
where
$ A(x) = \sqrt{\lambda}\> {\rm Ai}(x)$,
and the {\it sine kernel\/}
\[ {\lambda\over \pi}\> {\sin\pi(x-y)\over x-y} \, ,  \]
whose associated Fredholm determinant describes the
classical level spacing distribution  functions first studied
by Wigner, Dyson, Mehta, and others \cite{mehta_book}.
(In retrospect we should not have been surprised by this;
the two kernels are, after all, both scaled limits of the
same family of kernels.)
We describe below  three of these which we have found.
The first is the analogue  of the completely integrable system of
P.D.E.'s of Jimbo, Miwa, M{\^o}ri, and Sato~\cite{jmms}
 when  the underlying
domain is a union of intervals.
  The second is the fact that in
the case of the semi-infinite interval $(s,\infty)$ (the analogue
of a single finite interval for the sine kernel)
 the Fredholm determinant
is closely related to a Painlev{\'e} transcendent of the second
kind (the fifth transcendent arises for the sine kernel~\cite{jmms}).
  And the
third is the existence of a  second order differential
operator commuting  with the Airy operator $K$.  (The existence of such
a differential operator in the sine kernel case  has been known
for some time~\cite{meixner,fuchs_slepian}.)\    This last fact  leads to
an explicit asymptotic formula, as
the interval $(s,\infty)$ expands,  for the probability that it
contains precisely $n$ eigenvalues ($n=1,2,\ldots$)  (the analogue  of
results in \cite{btw_w}).
\section{ The system of P.D.E.'s}
\label{sec:pde}
We set
\[ I=\bigcup_{j=1}^m\left(a_{2j-1},a_{2j}\right)  \]
and write $D(I;\lambda)$ for the Fredholm determinant of $K$ acting
on $I$.  We think of this as a function of $a=(a_1,\ldots,a_{2m})$.
Then
\begin{equation}
 d_a \log D(I;\lambda) = - \sum_{j=1}^{2m} (-1)^j R(a_j,a_j)\, da_j
\label{dlog_tau}
\end{equation}
where $R(x,y)$ is the kernel of the operator $K(1-K)^{-1}$.
We introduce quantities
\[ q_j=(1-K)^{-1} A(a_j)\, , \ \ p_j=(1-K)^{-1} A^\prime(a_j)\, , \]
(which are the analogue  of the
quantities $r_{\pm j}$ of \cite{jmms};  see also
\cite{its_etal}) as well as two further quantities
\[ u=\left(A,(1-K)^{-1}A\right)\, , \ \ v=\left(A,(1-K)^{-1}A^\prime\right) \]
where the inner products refer to the domain $I$.  Then the equations read
\begin{eqnarray*}
{\partial q_j\over\partial a_k}&=& (-1)^k\,  {q_j p_k - p_j q_k\over
a_j-a_k}\, q_k\,\ \ \ (j\neq k),  \\
{\partial p_j\over\partial a_k}&=& (-1)^k\,  {q_j p_k - p_j q_k\over
a_j-a_k}\, p_k \ \ \  (j\neq k),  \\
{\partial q_j\over\partial a_j}&=&
-\sum_{k\neq j} (-1)^k\,  {q_j p_k - p_j q_k \over a_j - a_k}\, q_k
+p_j - q_j u\, , \\
{\partial p_j\over\partial a_j}&=&
-\sum_{k\neq j} (-1)^k\,  {q_j p_k - p_j q_k \over a_j - a_k}\, p_k
+a_j q_j + p_j u-2 q_j v\, , \\
{\partial u\over\partial  a_j}&=&(-1)^j q_j^2\, ,   \\
{\partial v\over\partial  a_j}&=& (-1)^j p_j q_j \, .
\end{eqnarray*}
Moreover the quantities $R(a_j,a_j)$ appearing in (\ref{dlog_tau})
are given by
\[ R(a_j,a_j)=\sum_{k\neq j}(-1)^k\,  {(q_j p_k - p_j q_k)^2\over a_j - a_k} \,
+ p_j^2 - a_j q_j^2 - 2 p_j q_j u + 2 q_j^2 v \, .\]
These equations are derived very much in the spirit of
 \cite{tracy_widom1}; see also \cite{its_etal}.
\section{The O.D.E.'s}
\label{sec:ode}
  For the special case $I=(s,\infty)$
the above equations
 can be used
to show that $q(s;\lambda)$ (the quantity $q$ of the last section corresponding
to the end-point $s$)
 satisfies
\begin{equation}
q^{\prime\prime}= s\,  q + 2 q^3 \, ,\ \ \ (^\prime={d\over ds})
\label{p2}
\end{equation}
with $ q(s;\lambda) \sim \sqrt{\lambda}\> {\rm Ai}(s)$
 as $s\rightarrow\infty$.  This equation
is a special case of the ${\rm P}_{II}$ differential equation
\cite{painleve,ablowitz_segur,mccoy_tracy_wu,hastings}.
One can similarly derive for $R(s):=R(s,s)$, which in view of
(\ref{dlog_tau}) equals
\[ {d\over ds}\, \log D(I;\lambda)\, , \]
the third-order equation
\begin{equation}
{1\over 2}\left( {R^{\prime\prime}\over R^\prime}\right)^\prime
- {R\over R^\prime} + R^\prime = 0\, .
\label{rDE}
\end{equation}
It is also the case that
$  R^\prime(s)=-q(s;\lambda)^2$ and this gives the following simple
formula for $D(I;\lambda)$
 in terms of a ${\rm P}_{II}$ transcendent:
\[ D(I;\lambda)  = \exp\left( -  \int_s^\infty  (x-s) q(x;\lambda)^2
\right)  \, dx \, . \]
This is much simpler than the corresponding representation of
$D(I;\lambda)$
for the sine kernel in terms of a ${\rm P}_V$ transcendent.
The fact that $q(s;\lambda)$ satisfies (\ref{p2}) can also be
obtained by combining some results in \cite{ablowitz_segur,hastings}.
Thus in this  case of a semi-infinite
interval  our results  have inverse
scattering interpretations.
\section{Asymptotics and Commuting Differential Operators}
\label{sec:commuting}
Again we take $I=(s,\infty)$ and consider asymptotics as $s\rightarrow
-\infty$. (Asymptotics as $s\rightarrow\infty$ can be obtained
trivially from the Neumann series for $(1-K)^{-1}$.)\   From the
random matrix point of view the interesting quantities are
\[ E(n;s):= {(-1)^n\over n!}\> {\partial^n\over \partial \lambda^n}
\, D(I;\lambda)\biggr\vert_{\lambda=1} \, .\]
This is the probability that exactly $n$ eigenvalues lie in $I$.
Using both differential equations (\ref{p2}) and (\ref{rDE}),
plus the fact that $R^\prime=-q^2$, we can obtain the asymptotics
of $R$ as $s\rightarrow -\infty$:
\[ R(s)\sim  {1\over 4} s^2 -{1\over 8\, s} + {9\over 64\,  s^4}
 -{189\over 128\,  s^7}  +\cdots \, . \]
(We also  use the  fact \cite{hastings}
that $q(s;1)$
is asymptotic to
$\sqrt{-s/2}$ as $s\rightarrow - \infty$.)\ \   Therefore
as $s\rightarrow -\infty$
\[ E(0;s) = D(I;1) \sim  {\tau_0\over (-s)^{1/8}}\exp\left(s^3/12\right)
\left (1 -
{3\over 64\,  s^3} +{2025\over 8192\, s^6}  + \cdots \right)
\]
where $\tau_0$ is an undetermined constant.  (The analogue  of this
formula for  the sine kernel  was  obtained by
Dyson \cite{dyson76}.)
\par
For asymptotics of $E(n;s)$ for general $n$ we introduce
\[ r(n;s):= {E(n;s)\over E(0;s)}\, . \]
Successive differentiation of (\ref{p2}) with respect to $\lambda$,
plus the known asymptotics of $q(s;1)$, allows us to find asymptotic
expansions for the quantities
\[ q_n(s):= {\partial^n q\over \partial\lambda^n}\biggr\vert_{\lambda=1}\]
(for the analogue  in the sine kernel case see \cite{btw_w}); and these
in turn can be used to find expansions for $r(n;s)$.  One drawback
of this approach is that yet another undetermined constant factor
enters the picture (in \cite{btw_w} Toeplitz and Wiener-Hopf techniques,
not available for the Airy kernel, fixed this constant).  Another
drawback is that when one expresses the $r(n;s)$ in terms of the
$q_n(s)$ a large amount of cancellation takes place, with the result
that even the first-order asymptotics of $r(n;s)$ are out of reach
by this method when $n$ is large.
\par
There is, however, another approach (briefly indicated in \cite{btw_w}
and with details in \cite{tracy_widom1} for the sine kernel case).
We have
\begin{equation}
r(n;s)=\sum_{i_1<\cdots<i_n} {\lambda_{i_1}\cdots \lambda_{i_n} \over
(1-\lambda_{i_1})\cdots (1-\lambda_{i_n}) }
\label{eig_repr}
\end{equation}
where $\lambda_0>\lambda_1>\cdots$ are the eigenvalues of the
integral operator $K$ (with $\lambda=1$).  Now quite analogous to
the fact that the operator with the sine kernel commutes with
the differential operator for the prolate spheroidal wave
functions \cite{meixner,fuchs_slepian}, is that  the Airy operator commutes
with the differential operator $L$ given by
\[ Lf(x)=\left(\left(x-s\right) f^\prime(x)\right)^\prime - x(x-s) f(x)\, .
\]
An application of the WKB method, plus a trick, allows us to
derive the following asymptotic formula for $\lambda_i$ with
$i$ fixed:
\[ 1-\lambda_i\sim {\sqrt{\pi}\over i!}\,  2^{5i+3}\,  t^{3i/2 +3/4} \exp\left(
-{8\over 3} t^{3/2}\right) \ \ \ (s=-2t\rightarrow - \infty).\]
(The analogue of this for the sine kernel is in~\cite{fuchs_slepian}.)
{}From this it is seen that the term in (\ref{eig_repr}) corresponding
to $i_1=0$, $i_1=1$, \ldots, $i_n=n-1$ dominates each of the others.
In fact this term dominates the sum of all the others, and so
\begin{equation}
r(n,-2t)\sim {1! 2! \cdots (n-1)!\over \pi^{n/2}\,  2^{(5n^2+n)/2}}\>
t^{-3n^2/4}\>\exp\left( {8\over 3} n t^{3/2} \right ) \, .
\label{r_asy}
\end{equation}
Thus one can use (\ref{r_asy}) to fix the constant in $q_1(s)$
mentioned above.

\acknowledgments
We wish to thank E.~Br{\'e}zin and P.~J.~Forrester for helpful
comments.  This work was supported in part by the National Science
Foundation, DMS--9001794 and DMS--9216203, and this support is
gratefully acknowledged.

\end{document}